\documentclass[aps,prl,twocolumn]{revtex4-1}

\usepackage[utf8]{inputenc}
\usepackage[english]{babel}

\usepackage{ulem}
\usepackage[colorlinks]{hyperref}
\usepackage{color}
\usepackage{mathtools}
\usepackage{latexsym}
\usepackage{graphicx}
\usepackage{float}
\usepackage{dcolumn}
\usepackage{bm}
\usepackage{amssymb}
\usepackage{amsmath}
\usepackage{calc}
\usepackage{mathtools}
\usepackage[space]{grffile}
\usepackage{xcolor}
\usepackage{comment}
\usepackage{epigraph}
\usepackage[makeroom]{cancel}

\usepackage{physics}

\definecolor{armygreen}{rgb}{0.29, 0.35, 0.09}

\newcommand{\pd}{{\phantom{\dagger}}}

\begin{document}

\title{Relaxation of phonons in the Lieb-Liniger gas  by dynamical refermionization }
\author{Isabelle Bouchoule$^{(1)}$}
\author{J\'er\^ome Dubail$^{(2)}$}
\author{L\'ea Dubois$^{(1)}$}
\author{Dimitri M. Gangardt$^{(3)}$}
\affiliation{$^{(1)}$Laboratoire Charles Fabry, Institut d’Optique Graduate School, CNRS, Université Paris-Saclay, 91127 Palaiseau, France\\
$^{(2)}$Universit\'e de Lorraine, CNRS, LPCT, F-54000 Nancy, France\\
$^{(3)}$School of Physics and Astronomy, University of Birmingham, Edgbaston, Birmingham, B15 2TT, UK
}
\date{\today}

\begin{abstract}
Motivated by recent experiments, we investigate the Lieb-Liniger 
gas initially prepared in an out-of-equilibrium state that is Gaussian in terms of the phonons,
namely whose density matrix is the exponential of an operator 
quadratic in terms of phonons creation and anihilation 
operators. 
Because the phonons are not exact eigenstates 
of the Hamiltonian, the gas relaxes to a stationary state at very long times whose phonon population is {\it a priori} different from the initial one. Thanks to integrability, that stationary state needs not be a thermal state. 
Using the Bethe Ansatz mapping  between the exact eigenstates of the Lieb-Liniger Hamiltonian and those of a non-interacting Fermi gas  and  bosonization techniques  we completely characterize the stationary state of the gas after relaxation and compute its phonon population distribution.
We apply our results to the case where the initial state is an excited coherent
state for a single phonon mode, and we compare them to exact results  obtained in the hard-core limit.

\end{abstract}

\maketitle


\paragraph{Introduction.}

Phonons are a central concept in the field of quantum gases. They are quantized sound waves, or collective phase-density excitations, that arise in low-energy and long wave-length  description of
quantum gases, e.g. in Bogoliubov theory of Bose-Einstein condensates in $D \geq 2$ spatial dimensions~\cite{pitaevskii2016bose}, or, in 1D, in Bogoliubov theory for quasi-condensates~\cite{popov_functional_1987,mora_extension_2003} and more generally in Luttinger liquid theory~\cite{haldane1981luttinger,cazalilla_bosonizing_2004,cazalilla_one_2011}. 
Phonons are routinely used to analyze experiments with out-of-equilibrium quantum gases, such as
the dynamics generated by a quench of the interaction strength in a 1D Bose gas~\cite{schemmer_monitoring_2018}, 
or by its  splitting into two parallel clouds~\cite{langen_local_2013,rauer_recurrences_2018}, or by a 
quench of the external potential~\cite{cataldini_emergent_2021}. 
In general, the description in terms of phonons accounts remarkably well for the 
observed short time dynamics~\cite{rauer_recurrences_2018,geiger_local_2014}. 
Crucially, the out-of-equilibrium states produced in these experimental setups are phononic Gaussian states with expectation values of phononic operators obeying Wick's theorem.
This is because they are obtained from a thermal equilibrium state, which is itself described by a Gaussian density matrix, by 
acting on it with linear or quadratic combinations of density and/or phase field.

However, although phonons  are exact eigenstates of the effective low-energy Hamiltonian, they are only approximations of the true eigenstates of the microscopic Hamiltonian. Thus, phonons have finite lifetime~\cite{Andreev1980,Samokhin1998,ristivojevic_decay_2014,ristivojevic_decay_2016,micheli_phonon_2022} and, at long times, the phonon distribution should evolve. In ergodic systems, this evolution would consist in the relaxation 
towards thermal equilibrium. 
But what if the microscopic system is integrable?
In this paper, we investigate the integrable Lieb-Liniger model of 1D Bosons with contact repulsive interactions
\cite{lieb_exact_1963-1}. 
 We express the phonons in terms of the true, {\it i.e.} infinite-lifetime, quasi-particles of the Lieb-Liniger model. We can then characterize the final stationary state of the system after relaxation and we relate the phonon mode occupations to the ones in the initial state.

\paragraph{Sketch of the main result.} The Hamiltonian of the Lieb-Liniger model is
\begin{equation}
    \label{eq:H_LL}
    H = \int_0^L dx~ \Psi^\dagger \left( - \frac{1}{2} \partial_x^2  + \frac{c}{2} \Psi^\dagger \Psi \right) \Psi ,
\end{equation}
with second-quantized bosonic operators obeying commutation relations $[\Psi(x) , \Psi^\dagger(y)]= \delta (x-y)$.
Here $c>0$ is the repulsion strength,  $L$ is the length of the system (we use periodic boundary conditions) and we use units such that  $\hbar =m=1$.
For $N$  atoms the average density is $\rho_0=N/L$.
The density fluctuation and current operators are
\begin{eqnarray}
    \label{eq:deltan_LL}
\nonumber    \delta \rho(x) &=& \Psi^\dagger(x) \Psi(x) - \rho_0 , \\
    J(x) &=& \frac{i}{2} \left[ (\partial_x \Psi^\dagger(x)) \Psi(x) - \Psi^\dagger(x)( \partial_x \Psi(x))  \right] .
\end{eqnarray}

At low temperature, it is customary to think of low-energy and long wavelength excitations above the ground state as quantized sound waves (phonons) that move to the right or to the left at the sound velocity $v$. The chiral combinations 
\begin{equation}
    \label{eq:RphononLL}
    J_{\rm R/L}(x) =  \frac{1}{2} \left( v~\delta \rho (x) \pm J(x) \right)
\end{equation}
are  the currents carried by right-moving (R) or left-moving (L) quasi-particles, with Fourier modes
\begin{align}
    \label{eq:aop}
    J_{\rm R}(x)=\frac{v\sqrt{K}}{L}\sum_{n>0}\sqrt{n} \left(e^{i \frac{2\pi n x }{L}}A_{{\rm R},n}+e^{-i \frac{2\pi n x }{L}}A^\dagger_{{\rm R},n}\right),
\end{align}
with  $n>0$ (a similar definition holds for $A_{{\rm L},n}$). Here $K=\pi \rho_0/v$ is the Luttinger parameter. 
Acting with $A^\dagger_{\mathrm{R/L},n}$ on the ground state $\left| 0 \right>$, one generates R/L-phonons.  We stress that the excited states generated this way are only {\it approximations} of the true eigenstates of the Lieb-Liniger Hamiltonian (\ref{eq:H_LL}). The lifetime of phonons may be large, but it is not infinite, so the phonon population will evolve, until it ultimately reaches a stationary value, governed by the {\it true eigenstates} of the Lieb-Liniger model.

Our main result is a general formula  that relates the phonon population at infinite time to the one in the initial state. 
The latter
is  assumed to be  Gaussian in terms of phonons, such that correlation functions of products of operators $J_{\rm R/L}(x)$ reduces to sums of products of one- and two-point correlation functions by Wick's theorem.
In the special case of a translation-invariant initial state $\left< . \right>_{0}$, populated by R-phonons,  
parametrized by a single function $g(x)$ via 
~\footnote{The function $g$ must satisfy $g(x) = g^*(-x)$, where $*$ denotes complex conjugation, and it must have a logarithmic singularity at short distance such that $ e^{g(x)} \frac{2\pi x}{i L } \rightarrow 1$ $x\rightarrow 0^+$, see below for details.}
\begin{equation}
    \left< J_{\rm R}(x) J_{\rm R}(y) \right>_0  \, =\,  - \frac{\rho_0 v}{4\pi}  \partial_x^2 g(x-y) ,
\end{equation}
and $\expval{J_R(x)}_0=0$, our result is that the two-point function ultimately relaxes to
\begin{equation}
    \label{eq:result_intro}
    \left< J_{\rm R}(x) J_{\rm R}(y) \right>_{\infty} \, =\,  \frac{ \pi \rho_0 v}{L^2} \exp \left(  2g(x-y) \right) .
\end{equation}
Thus, the phonon population evolves,
unless the function $g(x)$ satisfies $\partial_x^2 g = -(4\pi^2/L^2) e^{2 g}$. One solution to this equation is the thermal distribution  at inverse temperature $\beta\ll L/v$, for which $g(x-y) = - \log \left( \frac{2 \beta v }{iL} \sinh \left( \frac{\pi (x-y)}{\beta v} \right) \right)$ \cite{giamarchi2003quantum}. So the thermally occupied  phonon modes will not evolve,  but  more general initial states  will show a relaxation phenomenon.

In the rest of this Letter we derive Eq.~(\ref{eq:result_intro}) and its generalisation to 
initial Gaussian phononic states 
not  necessarly translationally invariant. 
We compare our 
predictions to exact numerical results obtained in the hard-core (Tonks-Girardeau) limit for an state with a single phononic mode initially displaced. We conclude by discussing perspectives
for experimental observation of the evolution of phonon populations.

\paragraph{Eigenstates of the Lieb-Liniger model and Bethe fermions.} For even/odd $N$,
an $N$-particle eigenstate of (\ref{eq:H_LL}) is specified by an ordered set of half-integers/integers 
$I_1<I_2<\ldots<I_N$ which uniquely determines the set of $N$ rapidities $\lambda_1<\lambda_2<\ldots<\lambda_N$ via the Bethe equations~\cite{lieb_exact_1963-1,gaudin2014bethe,korepin1997quantum} 
\begin{equation}\label{eq:BA}
	\lambda_a  + \frac{1}{L} \sum_{b=1}^N {\rm arctan} \left( \frac{\lambda_a-\lambda_b}{c} \right) \, = \, \frac{2\pi}{L} I_a \, .
\end{equation}
The energy of that eigenstate is $E_{\{\lambda_a\}} = \sum_a \lambda_a^2/2$ and the corresponding wavefunction is
$
    \left< {\rm vacuum} \right| \prod_{j=1}^N \Psi(x_j)  \left| \{\lambda_a \}\right> \, \propto \sum_{\sigma} \mathcal{A}_\sigma  e^{i \sum_a \lambda_{\sigma(a)} x_a }
$, where the sum is over all permutations $\sigma$ of $N$ elements
and  $\mathcal{A}_\sigma = \prod_{a>b} \left( 1 - i c~{\rm sgn}(x_a-x_b)/(\lambda_{\sigma(a)}-\lambda_{\sigma(b)}) \right)$. In the following we assume $N$ even. The ground state corresponds to  densely packed Bethe half-integers
$	\{ I^{(0)}_a\} \, = \, \left\{ -\frac{N-1}{2} , -\frac{N-1}{2}+1, \dots , \frac{N-1}{2} \right\} $.

Eq. (\ref{eq:BA}) provides a one-to-one mapping between the eigenstates $\ket{\{I_a\}}$ of the Lieb-Liniger model  and the eigenstates of $N$ non-interacting fermions with momenta $2\pi I_a/L$, $a=1,\ldots,N$. This mapping preserves the total momentum $P_{\{\lambda_a\}} = \sum_a \lambda_a = (2\pi/L)\sum_a I_a$. It is natural to introduce fermion operators $b_J$ that act on the normalized eigenstate $\ket{\{I_a\}}$ by removing a Bethe half-integer $J$ from the set $\{I_a\}$, if it is present, and by annihilating the state otherwise. Conversely, the operator $b^\dagger_J$ inserts $J$ in the set $\{I_a\}$ unless it is already present. The eigenstate corresponding to the modified state is then multiplied by $(-1)^{n_{I_a<J}}$, where $n_{I_a<J}$ is the number of elements of $\{I_a\}$ 
smaller than $J$, to enforce the correct anti-commutation relations for the `Bethe fermion' operators $b_J^\dagger/b_J$ 
~\footnote{Acting with an odd number of such fermion operators changes parity of $N$, thus changing 
the boundary conditions for the bosons. We can ignore this here, because we restrict to excitations that conserve the atom number.}.

All eigenstates of (\ref{eq:H_LL}) with a total atom number $N$ are generated by acting on the ground state with an equal number of Bethe fermion creation and annihilation operators.
In particular, the low energy states 
are obtained by acting with creation/annihilation operators close to the R/L Fermi points. For a half-integer $l$, we define the operators
\begin{equation}
	c_{{\rm R},l}^\dagger = b^\dagger_{\frac{N}{2} + l} , \qquad  c_{{\rm L},l}^\dagger = b^\dagger_{ -\frac{N}{2} -l} .
\end{equation}
The low energy eigenstates $\ket{\psi}$ with $q$ R-excitations 
are of the form 
\begin{equation}
    \label{eq:low_energy}
\ket{\psi} =  \prod_{i=1}^q  c^\dagger_{\mathrm{R}, l_i } \prod_{j=1}^q c_{\mathrm{R}, m_j }  \ket{0} ,
\end{equation}
for sets of half-integers $l_j > 0$ and $m_j < 0$, with $q \ll N$, and $|l_j | , |m_j| \ll N$. 
To lighten our formulas, 
we consider eigenstates with R-excitations only; 
it is straightforward to generalize our results to include also L-excitations. 
The energy of the low-energy eigenstate (\ref{eq:low_energy}) is $E= \sum_{j=1}^q (\epsilon(l_j)-\epsilon(m_j))$ with  the `dressed' energy~\cite{Lieb_exact_1963} 
given approximately by the quadratic dispersion relation $\epsilon(l) = \frac{2\pi v l}{L} + \frac{1}{2m^*}\left(\frac{2\pi l}{L}\right)^2+O(1/L^3)$ with the effective mass $m^*=(1+(\rho_0/v)\partial v/\partial \rho_0)$ \cite{rozhkov_fermionic_2005,Pereira2006,Imambekov2012}.

\paragraph{Phonons.} 
\label{sec:bosons} We will make extensive use of the following simple formula for the matrix elements of $A^\dagger_{\mathrm{R},n}$ (Eq.~(\ref{eq:aop})) between two low-energy states of the form (\ref{eq:low_energy}) in the thermodynamic limit,
 \begin{equation}
    \label{eq:phononscreationOp_R}
     \left< \psi_2 \right| A_{{\rm R},n}^\dagger \left|\psi_1 \right>  \underset{N \rightarrow \infty}{=}   \frac{1}{\sqrt{n}} \left< \psi_2 \right| \sum_{l \in \mathbb{Z}+\frac{1}{2}} c^\dagger_{\mathrm{R}, n+l} c^\pd_{\mathrm{R},l} \left| \psi_1 \right>.
 \end{equation}
  Eq.~(\ref{eq:phononscreationOp_R}) follows from known results about form factors of the density operator in the Lieb-Liniger model, see Refs.~\cite{kozlowski2011form,kozlowski2015microscopic,de2015density,de2018edge} and~\cite{SM}. It shows that a phonon created by $A^\dagger_{{\rm R},n}$ is a coherent superposition of Bethe fermion particle-hole pairs
   and 
 that  phonons 
are obtained by bosonization of the Bethe fermions~\cite{giamarchi2003quantum,gogolin_bosonization_2004}.
 This implies that $A_{R,n}$ and $A_{R,n}^\dagger$ satisfy bosonic
canonical commutation rules \cite{giamarchi2003quantum,gogolin_bosonization_2004}
\begin{align}
    \label{eq:bos_comm}
[A^\pd_{{\rm R},n},A^\dagger_{{\rm R},n'}] = \delta_{n,n'} . 
\end{align}

Bosonization 
allows to invert Eq.~\eqref{eq:phononscreationOp_R} and represent the Bethe fermion operators $c^\dagger_{\rm R}(x) =  \sum_l e^{-i \frac{2\pi l x}{L}} c^\dagger_{{\rm R},l}/\sqrt{L} $ as
\begin{equation}
    \langle \psi_2| c_{\rm R}^\dagger(x)c_{\rm R}(y)|\psi_1\rangle= \langle \psi_2|:e^{-i\varphi_{\rm R}(x)}::e^{i\varphi_{\rm R}(y)}:|\psi_1\rangle/L,
\label{eq:bosonization}
\end{equation}
where the notation $:.:$ denotes normal ordering and $\varphi_{\rm R}(x)\!=\!-i\sum_{n>0}  (e^{i2\pi n x/L}A^\dagger_{R,n}-e^{-i2\pi n x/L}A^\pd_{R,n})/\sqrt{n}$ is the chiral field, related to the chiral current by  $J_{\rm R}(x)=v\sqrt{K}/(2\pi) \partial_x\varphi_{\rm R}(x).$
The bosonization formulas require  that
$\langle \varphi_{\rm R}(x)\varphi_{\rm R}(y)\rangle$
has the same short-distance logarithmic divergence
as the one in the ground state, $\langle \varphi_{\rm R}(x)\varphi_{\rm R}(y)\rangle =
-\log(2\pi (x-y+i\epsilon)/iL)$ as $y \rightarrow x$~
\footnote{The regulator $\epsilon \rightarrow  0^+$  ensures convergence of the sum in (\ref{eq:aop}). 
For $\varphi_{\rm L}(x)$, the result is similar, with $\epsilon \rightarrow - \epsilon$.},  which implies that the phonon population $\langle A_{{\rm R}, n}^\dagger A_{{\rm R}, n} \rangle$ decays at least exponentially with $n$.

\paragraph{Initial state preparation and short time dynamics.}
\label{sec:initialstate}
For short times the 
nonlinearity of the fermionic spectrum 
has small 
effect and, as one restricts to 
low-energy and long wavelength states, one can approximate the Lieb-Liniger Hamiltonian, Eq.~(\ref{eq:H_LL}), by the Luttinger Liquid Hamiltonian 
\begin{equation}\label{eq:Luttinger}
     H 
     \simeq v \sum_{n>0} 
      \frac{2\pi n}{L} \left ( A^\dagger_{{\rm R},n}A^\pd_{{\rm R},n}+A^\dagger_{{\rm L},n}A^\pd_{{\rm L},n}\right )\, .
\end{equation}
This Hamiltonian permits efficient calculation
of equal-time correlation functions at thermal equilibrium 
~\cite{mathey_noise_2009,imambekov_density_2009}. 
As explained in the introduction, it also describes successfully 
several experiments 
probing out-of-equilibrium dynamics
\cite{schemmer_monitoring_2018,langen_local_2013,rauer_recurrences_2018}.
In those 
experiments, the initial state is Gaussian in terms of phononic operators 
which motivates our choice to consider 
an initial phononic Gaussian state. The latter is characterized by the one- and (connected) two-point correlations functions of the chiral currents, that we parameterize in terms of functions $f(x)$ and $g(x,y)$ as
\begin{align}\label{eq:initialstate1}
	\left< J_{\rm R} (x) \right>_{0} &= v\frac{\sqrt{K}}{2\pi} \partial_x f_{\rm R} (x) \\  \left< J_{\rm R} (x) J_{\rm R}(y) \right>^{\rm conn.}_{0} =&
	v^2\frac{K}{(2\pi)^2} \partial_x \partial_y g_{{\rm RR}} (x,y) ,
	\label{eq:initialstate2}
\end{align}
and similarly for $\left< J_{\rm L} (x) \right>_{0}$ and $ \left< J_{\rm L} (x) J_{\rm L}(y) \right>^{\rm conn.}_{0} $, as well as for the possible cross correlation $\left< J_{\rm R} (x) J_{\rm L}(y) \right>^{\rm conn.}_{0} $. 
Here 
$\left< J_{\rm R} (x) J_{\rm R}(y) \right>^{\rm conn.} =\left< J_{\rm R} (x) J_{\rm R}(y) \right>-\left< J_{\rm R} (x)\right > \left < J_{\rm R}(y) \right>$.
Higher order correlation functions 
    are obtained from those   by Wick's theorem for the phononic operators.

\paragraph{Long time dynamics and relaxation.}
 The key point of 
 this paper is that the phononic  states are not eigenstates of
 the Lieb-Liniger Hamiltonian and therefore are not well adapted to study  the long time evolution. 
 This is clearly seen by examining the phase difference accumulated between different particle-hole states entering 
a phononic excitation given by the r.h.s. of Eq.~\eqref{eq:phononscreationOp_R}: one can estimate the relevant time scale for the 
dephasing  of a single phonon with momentum $n$ as
 $   t_\mathrm{deph} = \hbar m^* \left(\frac{L}{2\pi n}\right)^2.$

  The long-time behavior of the Lieb-Liniger gas is now 
  well established~\cite{caux_constructing_2012,palmai_quasilocal_2018}.
The system shows a relaxation phenomena:
 as long as local observables are concerned, the density matrix at long times
 is obtained from the initial one 
 by retaining only its diagonal elements
 in the Bethe-Ansatz eigenbasis. 
Moreover, according to the 
 Generalized Eigenstate Thermalization Hypothesis~ \cite{caux2013time,cassidy2011generalized,d2016quantum}
 which states that
all eigenstates 
are locally identical  
provided they have the same coarse grained rapidity distribution 
$\rho(\lambda)=(1/L)\sum_a \delta(\lambda-\lambda_a)$,
all diagonal density matrix sufficiently peaked around the correct rapidity distribution~\cite{caux2013time,cassidy2011generalized,d2016quantum}
are acceptable.
In this paper, we choose the Gaussian density matrix~\cite{rigol2007relaxation,vidmar2016generalized}
\begin{equation}
	\hat{\rho}_{\infty} \, \propto \, {\rm exp} \left( \sum_{I}  \beta^\pd_I b^\dagger_I b^\pd_I   \right )
	\label{eq:GGEBI}
\end{equation}
where the distribution of Bethe half-integers imposed by the Lagrange multipliers $\beta^\pd_I $ ensures the 
correct distribution of rapidities~\footnote{The exact 
density matrix is $\hat{\rho}_{\rm GGE} \, \propto \,  \Pi_{o}e^{ \sum_{I\in \mathbf{Z}}  \lambda_I b^\dagger_I b_I  } + \Pi_{e}e^{\sum_{I\in \mathbf{Z} +1/2}  \lambda_I b^\dagger_I b_I  } $, where $\Pi_o$, resp. $\Pi_e$, are the projectors on the states with odd, resp. even, atom number. Since we consider operators that conserve atom number, 
the expression
given 
in Eq.~\eqref{eq:GGEBI} is sufficient.}.
A commonly used  
 alternative is 
 the Generalized Gibbs Ensemble in terms of the  rapidity distribution. 
Both ensembles are equally valid
as long as local quantities are concerned~\footnote{ 
The relevance of the GGE in terms of rapidities shows when
it is applied to describe a sub-system of size $l$, 
much smaller than the total system size, but much larger than microscopic lengths: it will gives correct mean values 
for operators acting on the sub-system, even non-local ones, 
up to corrections of the order $1/l$.}.

\begin{figure*}
    \centerline{\includegraphics[width=1\linewidth]{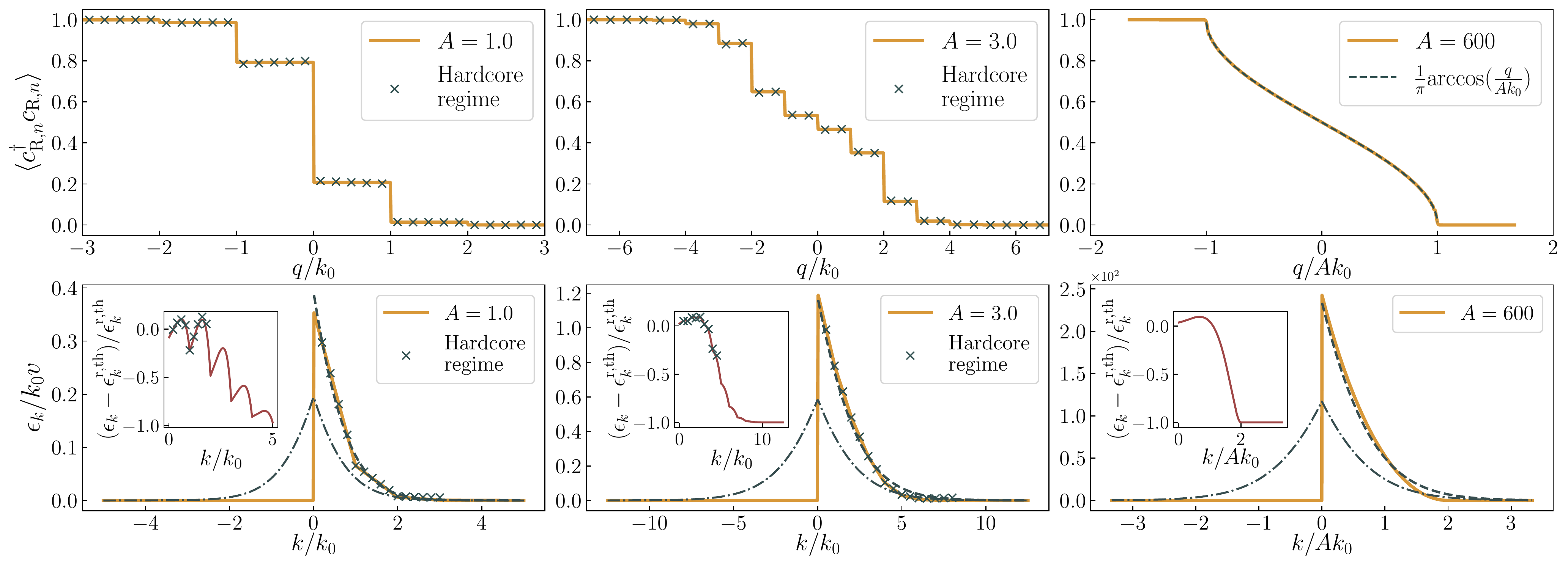}}
    \caption{Relaxation of a single displaced phononic mode,
    characterized by  $f_R(x)=A\cos(k_0 x)$. 
    Solid yellow lines are obtained by dynamical re-fermionization, {\it i.e.} Eqs.~(\ref{eq:gr}-\ref{eq:Grversuscdagxcy}-\ref{eq:cdagxcyversusgf}-\ref{eq:phononpoprelaxcndagcn}), in the thermodynamic limit $L\rightarrow \infty$.  
    Top: Bethe-fermions occupations in the vicinity of  the Fermi sea right-border.
      Bottom: Energy in each phononic mode  after relaxation, $ \epsilon_{k}= |k| v 
      \langle A^\dagger_{R,n}A_{R,n}\rangle_{\infty}$, where $k= 2n\pi/L$ (resp. $-2n\pi/L$) for right-movers (resp. left-movers). The dashed-dotted black lines 
      is the results expected if the system would relax to 
      a  thermal equilibrium. 
     Dashed line shows the distribution $\epsilon_k^{\rm r,th}$ corresponding to thermal  redistribution of the initial energy within the right-movers phonons only and the insets highlight its difference with the 
     dynamical refermionization results.  
     Crosses are exact numerical 
     results obtained in the limiting case of 
     hard-core bosons, performed for 100 atoms (see text), using 
     $k_0 = 5 \times 2 \pi /L$ in the left figure (resp. $k_0  = 2 \times 2 \pi /L$ in the central figure) for $A = 1.0$  (resp. $A = 3.0$), and the relaxed phonon population is computed in the fermionic diagonal ensemble. Those numerical results are performed for a lattice gas of 1000 sites, which represents faithfully the  continuous gas only for small $k$.}
    \label{fig:populations}

\end{figure*}

Extracting the numbers $\beta_I$, or equivalently the expectations $\langle b^\dagger_Ib^\pd_I\rangle$ from the 
correlation functions Eqs.~\eqref{eq:initialstate1}-\eqref{eq:initialstate2}
which parameterize the initial phononic Gaussian state, is,
generally speaking, an excruciating task. However, for an initial state
in the low energy sector, 
only fermionic states  
close to the Fermi points are affected and calculation of 
$\langle b^\dagger_Ib^\pd_I\rangle$, which
reduces to finding the  distributions $\langle c^\dagger_{{\rm R},n}c^\pd_{{\rm R},n}\rangle$, $\langle c^\dagger_{{\rm L},n}c^\pd_{{\rm L},n}\rangle$,  is much an easier task as it
can be done by using
bosonization.

To do this we concentrate on the right movers and introduce $G_\mathrm{R}(\xi)$ defined by 
\begin{align}\label{eq:gr}
   G_\mathrm{R}(\xi) = \frac{1}{L}\sum_l e^{-i2\pi l\xi/L}\langle c^\dagger_{\mathrm{R},l}c^\pd_{\mathrm{R},l}\rangle\, .
\end{align}
which can be rewritten as the spatially averaged fermion two-point correlation function
\begin{align}
\label{eq:Grversuscdagxcy}
    G_\mathrm{R}(\xi) 
=(1/L)\int du \langle c^\dagger_{{\rm R}} (u+\xi/2) c^\pd_{{\rm R}}(u-\xi/2) \rangle\, .
\end{align}

The crucial observation is that since $G_R$ is time-independent it can be evaluated using the initial state. Since the latter is 
a phononic Gaussian state,
one can use Wick's theorem for 
$\varphi_{\rm R}$ in Eq.\eqref{eq:bosonization} to evaluate of the two-point fermionic correlation function.  One obtains, 
using $\langle \varphi_{\rm R}(x)\rangle_0=f_{\rm R}(x)$
and $\langle \varphi_{\rm R}(x)\varphi_{\rm R}(y)\rangle^{\rm conn.}_0=g_{\rm RR}(x,y)$,
\begin{align}
\label{eq:cdagxcyversusgf}
    &\langle c^\dagger_{{\rm R}}(x) c^\pd_{{\rm R}}(y)\rangle_{0}   = \frac{1}{L} 
    \exp\left[-i  (f_{{\rm R}} \left(x\right) -  f_{{\rm R}}\left(y\right)) \right]\notag \\
    &\times
   \exp\left[g_{\rm RR} \left(x ,y \right)
        - \frac{1}{2} g_{\rm RR}^{\rm reg.} \left(x\right)  -\frac{1}{2} g_{\rm RR}^{\rm reg.} \left(y\right)\right] 
\end{align}
where
$g_{{\rm RR}}^{\rm reg.} (x)  \, = \, 
  \lim_{y \to x} g_{{\rm RR}} (x, y) + \log (2\pi(y-x)/(iL))$
  is independent of  the short distance cutoff $\epsilon$. The function $G_R(\xi)$ is obtained by injecting Eq.~\eqref{eq:cdagxcyversusgf}
   into Eq.~\eqref{eq:Grversuscdagxcy}. The
   population of the Bethe fermions, which entirely characterizes the state after relaxation, is then computed  inverting Eq.~\eqref{eq:gr}. 
    We dub this crucial intermediate result  `dynamical 
     refermionization'.

\paragraph{Consequence: relaxation of phonon population.}
Mean values of products of phononic operators  after
relaxation are computed 
expressing them in terms of fermionic operators thanks to
Eq.\eqref{eq:phononscreationOp_R},
and using Wick's theorem, valid for the
fermionic Gaussian density matrix Eq.~\eqref{eq:GGEBI}.
In particular, to compute $\langle J_{\rm R}(x)J_{\rm R}(y)\rangle_\infty$, we use the relation   
$J_{\rm R}(x)=(v\sqrt{K}/L) \sum_{l}\sum_{n\neq 0} e^{i2\pi n x/L} c^\dagger_{{\rm R},l+n} c^\pd_{{\rm R},l}  $, 
obtained injecting Eq.~\eqref{eq:phononscreationOp_R} into \eqref{eq:aop}. 
This gives, for $x\neq y$,
\begin{equation}
\label{eq:JrxJryrelax}
  \langle J_{\rm R}(x)J_{\rm R}(y)\rangle_{\infty} = -K v^2 G_{\rm R}(x-y)G_{\rm R}(y-x) .
\end{equation}
Also $\expval{J_\mathrm{R}(x)}_\infty=0$ due to translational invariance. The phonon population reads
\begin{equation}
    \langle A^\dagger_{{\rm R},n}A_{{\rm R},n}\rangle_{\infty}=
    \frac{1}{n}\sum_{l} \langle c^\dagger_{{\rm R},l+n}c_{{\rm R},l+n}
    \rangle \left ( 1- \langle c^\dagger_{{\rm R},l}c_{{\rm R},l}
    \rangle \right ).
    \label{eq:phononpoprelaxcndagcn}
\end{equation}
 Note that one should 
consider its weighted sum 
over a small 
but non-vanishing width in $k=2\pi n/L$, to ensure that the quantity is 
local so Eq.~(\ref{eq:GGEBI}) applies.

Eqs.~\eqref{eq:JrxJryrelax}-\eqref{eq:phononpoprelaxcndagcn} constitute the main result of this paper. 
The translation-invariant case, Eq.~\eqref{eq:result_intro}, announced earlier, is obtained by using  $f_{\rm R} (x) = 0$, $g_{\rm RR} (x,y) = g (x-y)$, $g^\mathrm{reg.}_\mathrm{RR} (0) = 0$.
We stress that the relaxed  state of the  system is no longer Gaussian in terms of the phonons, so that higher order phononic correlation functions 
would require a separate calculation.

\paragraph{Example: application  to the case of a single excited phononic mode.} 
Let us consider the situation where the initial state is obtained from the ground state by a displacement 
of an R-phonon:  
 $f_{\rm R}(x)=A\cos(k_0 x)$ with the amplitude $A$  and the wave-vector  $k_0$, while keeping  
 $g_{\rm RR}(x,y)$  equal to its ground-state value. 
Fig.~\ref{fig:populations} shows the Bethe fermion distribution obtained from dynamical refermionization, Eqs.~(\ref{eq:gr}),(\ref{eq:Grversuscdagxcy}),(\ref{eq:cdagxcyversusgf}). 
For small amplitudes $A$, one observes plateaus of width $k_0$
which reflect the quantization of phonons~\cite{SM}.
As $A$ increases, 
more plateaus appear,
and for large $A$ it becomes the smooth profile $\langle c_{{\rm R},n}^\dagger c_{{\rm R},n}\rangle=(1/\pi){\rm arccos}(2\pi n/(LA k_0))$  expected semiclassically~\cite{SM}.
The bottom row of Fig.~\ref{fig:populations} shows the energy of each phononic mode after relaxation. 
The difference of the distributions of R and L-phonons 
is a strong signature of the non-thermal nature of the relaxed system.
Within the space of R-phonons, redistribution of energy among phonons is found to be  very efficient: the relaxed distribution is close, albeit not identical, to  that expected for a thermal state. 
We compare the dynamical refermionization predictions to exact results in the asymptotic regime of hard-core bosons ($c\rightarrow \infty$): 
in this regime, the 
Hamiltonian in term of Bethe fermions is that of non-interacting 
fermions and the current operator $J_R$ is 
equal to that for the Bethe fermions, which enables exact calculations. 
The initial state is obtained as the ground 
state of the Hamiltonian $H+(Ak_0/\sqrt{K})\int dx J_{\rm R}(x)\sin(k_0 x)$. 
As seen in Fig.~(\ref{fig:populations}), results are
in excellent agreement with the predictions of dynamical refermionization.

\paragraph{Experimental perspectives.}
Our predictions can be tested in  cold atom experiments, where
initial out-of-equilibrium states can be generated in various ways.
By quenching the longitudinal potential from a long-wavelength sinusoidal potential to a flat potential~\cite{cataldini_emergent_2021}, one produces
displaced phononic states, corresponding to non vanishing 
functions $f_{\rm R}$, $f_{\rm L}$. A quench of the interaction strength will produce
two-modes squeezed phononic states~\cite{schemmer_monitoring_2018}, a situation which 
corresponds to 
$f_{\rm R}=f_{\rm L}=0$, but to modified functions $g_{{\rm R}{\rm R}}$, $g_{{\rm L}{\rm L}}$, $g_{\rm RL}$. 
Alternatively, modulating the coupling constant 
with time
will parametrically excite only part of the phononic spectrum~\cite{jaskula_acoustic_2012}.
In the above scenarios the prepared initial state is symmetric under exchange of R and L-phonons. To break this symmetry, one could expose the gas to  a potential 
$V(x)=V_0\cos(k_0 x - vk t)$ for some short time duration: then only the R-phonons would be resonantly excited.

To probe the phonon distribution after relaxation, 
one possibility is to measure the in situ
long-wavelength density fluctuations~\cite{esteve_observations_2006} and access $\delta \rho (x) = (J_{\rm R}(x) + J_{\rm L}(x))/v$, see Eq.~\eqref{eq:deltan_LL}. Alternatively, one can use the density ripple techniques to 
probe the long wavelength phase fluctuations~\cite{manz_two-point_2010,schemmer_monitoring_2018}, 
whose gradient is the velocity field proportional to $J(x) = J_{\rm R}(x) - J_{\rm L}(x)$.
The above methods however do not discriminate between right and left movers. In order to probe selectively R-phonons, one needs 
to probe the dynamics, for instance using sequences of non-destructive  images~\cite{andrews_propagation_1997}. 

At very long times, 
one expects integrability breaking perturbations to bring the system to a thermal equilibirum. However, such perturbations can be weak enough to have negligible effect during the relaxation time of the Lieb-Liniger phonons, as is observed experimentally in Ref. \cite{cataldini_emergent_2021} and modeled recently in Ref.~\cite{moller2022bridging}.

Interestingly, the occupation of Bethe fermions $\langle c^\dagger_{{\rm R}, n} c_{{\rm R}, n} \rangle$ ---which, in this Letter, is used as an intermediate result 
--- could also be measured experimentally.
To measure it, one could first perform an adiabatic increase of the repulsion strength $c$, which preserves the distribution of Bethe fermions~\cite{bastianello_generalized_2019}, until the hard-core regime is reached. In this regime the distribution of Bethe fermions is the same as the rapidity distribution, which can be measured by a 1D expansion~\cite{wilson2020observation,bouchoule_generalized_2022}.

\paragraph{Prospects.}
This work calls for further 
investigations in several directions.  
First, one could investigate 
higher order functions of the chiral currents or of the phonons
populations to show that the relaxed state is non Gaussian with respect to the phonons.
Secondly, our predictions call for numerical studies of
relaxation in the Lieb-Liniger model away from the strongly interacting regime.
Finally, as discussed above, the predictions of this Letter are 
to be confirmed
experimentally.

\begin{acknowledgments}
We thank Jacopo de Nardis and Karol Kozlowski for very helpful discussions about Eq.~(\ref{eq:phononscreationOp_R}) and its relation to results on form factors in the literature. JD and DMG acknowledge hospitalit  y and support from  Galileo Galilei Institute, Florence, Italy, 
during the program "Randomness, Integrability, and Universality", where part of this work was done. This work was supported (IB-JD-LD) by the ANR Project QUADY - ANR-20-CE30-0017-01.
\end{acknowledgments}

\normalem
%

\newpage
\appendix
\section{Matrix element of density and current operators between two low-energy eigenstates of the Lieb-Liniger Hamiltonian}

\subsection{Density operator}

Various formulas are available in the literature for form factors of the density operator $\delta \rho(x)$, see e.g.~\cite{slavnov1989calculation,kozlowski2011form,kozlowski2011long,kozlowski2015microscopic,de2015density,de2018edge}. Here we use formula (8) of Ref.~\cite{de2018edge} as a starting point. That formula says that the matrix element between a Bethe state $\left| \psi \right>$ and another Bethe state $\left| \psi' \right>$, multiplied by the system size $L$, vanishes in the thermodynamic limit, unless $\left| \psi' \right>$ is obtained from $\left| \psi \right>$ by a single particle-hole excitation. In that case, if $p$ is the rapidity of the particle and $h$ is the rapidity of the hole, then the matrix element is
\begin{equation}
   L \left| \left< \psi' \right| \delta \rho(x) \left| \psi \right> \right| \underset{p\rightarrow h}{=} k'(p) \left| \frac{\lambda_0- h}{\lambda_0- p} \right|^{\Delta \vartheta (F(\lambda_0 | p ) - F(\lambda_0 | h ) )}
\end{equation}
where $k(\lambda)$ is the dressed momentum as a function of the rapidity $\lambda$, $\lambda_0$ is the rapidity at the Fermi point, $\Delta \vartheta$ is the height of the discontinuity of the occupation ratio at the Fermi point (here, for the ground state, we simply have $\Delta \vartheta= 1$), and $F(\lambda | \lambda' )$ is the backflow, see Ref.~\cite{de2018edge}. For low-energy excitations around the ground state, both $p$ and $h$ are very close to $\lambda_0$, so 
\begin{eqnarray*}
   L \left| \left< \psi' \right| \delta \rho(x) \left| \psi \right> \right|  & \underset{p, h \rightarrow \lambda_0 }{=} & k'(\lambda_0) e^{{\partial_{\lambda'} F(\lambda_0 | \lambda' )_{| \lambda' = \lambda_0}  \,  (p-h)   \log \left| \frac{\lambda_0- h}{\lambda_0- p} \right| } } \\
    &=& k'(\lambda_0) \left( 1+ O ( | p-h | \log |p-h|  ) \right),
\end{eqnarray*}
and the derivative of the dressed momentum at the Fermi rapidity is the square root of the Luttinger parameter, $k'(\lambda_0)  = \sqrt{K}$ (see e.g. Ref.~\cite{korepin1997quantum}). So, in the notations of our main text, we have that the matrix element of the Fourier mode $\delta \rho_n  = \int_0^L e^{i 2 \pi n x/L} \delta \rho(x) dx$ between two low-energy Bethe states $\left| \psi_1\right>$, $\left| \psi_2\right>$ in the thermodynamic limit is
\begin{eqnarray}
    \label{eq:matrix_elt_density}
    && \left< \psi_2 \right| \delta \rho_n \left| \psi_1 \right> = \\
 \nonumber    && \sqrt{K} \left< \psi_2 \right| \sum_{l \in \mathbb{Z} + \frac{1}{2}} c^\dagger_{{\rm R}, n + l} c_{{\rm R}, l} + \sum_{l \in \mathbb{Z} + \frac{1}{2}} c^\dagger_{{\rm L}, n + l} c_{{\rm L}, l} \left| \psi_1 \right>, 
\end{eqnarray}
up to corrections of order $O ( \log L / L) $.

\subsection{Current operator}

A similar formula is obtained for the expectation value of the Fourier modes of the current operator, $J_n = \int_0^L e^{i \frac{2\pi n x}{L}} J(x) dx$. It follows from the one for the density operator, and from continuity equation. Indeed, using
\begin{equation}
     \partial_x J(x)  = - \partial_t \delta \rho (x) = - i [H, \delta \rho(x)]
\end{equation}
one finds
\begin{eqnarray}
 \nonumber  - i \frac{2\pi n}{L} \left< \psi_2 \right| J_n \left| \psi_1 \right> &=& -i \left< \psi_2 \right| [H, \delta \rho_n] \left| \psi_1 \right> \\
   &=& -i (E_2 - E_1) \left< \psi_2 \right| \delta \rho_n \left| \psi_1 \right>,
\end{eqnarray}
where $E_1$ and $E_2$ are the energies of the eigenstates $\left| \psi_1 \right>$, $\left| \psi_2 \right>$. According to the expression given in the main text, the difference between energies for a single particle-hole excitation of momentum $p-h = \frac{2\pi n}{L}$ is $E_2 - E_1 = \epsilon(p) - \epsilon(h) \simeq  \frac{2\pi v n}{L}$. This, together with Eq.~(\ref{eq:matrix_elt_density}), gives the matrix element of the current operator in the thermodynamic limit,
\begin{eqnarray}
    \label{eq:matrix_elt_current}
    && \left< \psi_2 \right| J_n \left| \psi_1 \right> = \\
 \nonumber    && v \sqrt{K} \left< \psi_2 \right| \sum_{l \in \mathbb{Z} + \frac{1}{2}} c^\dagger_{{\rm R}, n + l} c_{{\rm R}, l} - \sum_{l \in \mathbb{Z} + \frac{1}{2}} c^\dagger_{{\rm L}, n + l} c_{{\rm L}, l} \left| \psi_1 \right>, 
\end{eqnarray}
up to corrections of order $O ( \log L / L) $.
The matrix elements (\ref{eq:matrix_elt_density}) and (\ref{eq:matrix_elt_current}) then lead to formula~(\ref{eq:phononscreationOp_R}) in the main text.

\section{Asymptotic behavior of the Bethe fermion distribution for a displaced phonon mode at $T=0$}

\begin{figure}[htb]
    \centering
    \includegraphics[width=\linewidth]{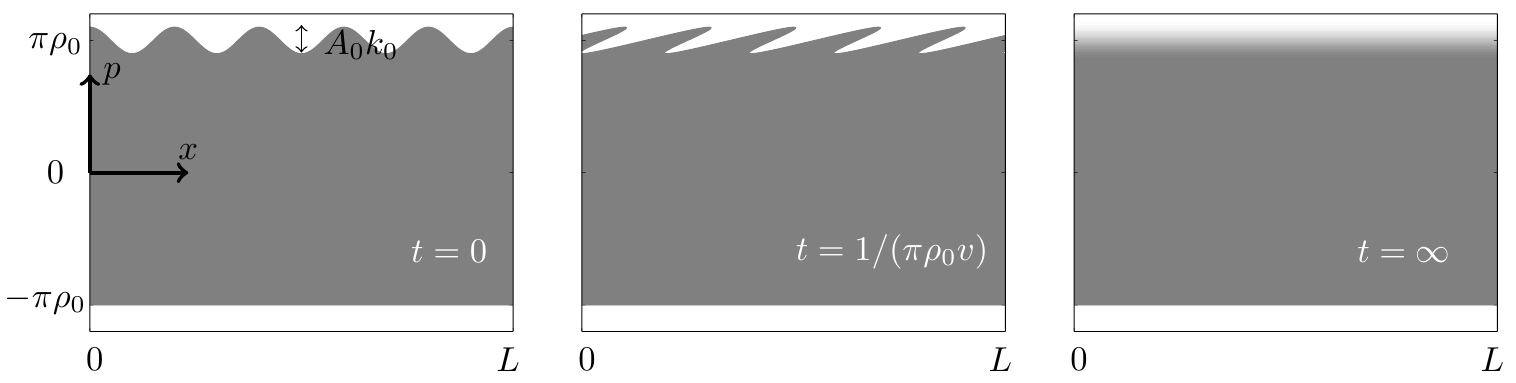}
    \caption{Illustration of dynamical refermionization in the semiclassical limit, when the number of phonons in the initial states is large. For simplicity we focus on the hard-core limit, where the dynamics of the bosons maps to the one of non-interacting fermions. Semiclassically, the fermions at zero temperature fill a region of phase space (grey area). In the initial state, this region is the set of all points $(x,p)$ with $0 \leq x < L$ and $\pi \rho_0 \leq p \leq  \pi \rho_0 - A_0 k_0 \sin (k_0 x)$. Then, at later times $t$, all points in phase space get translated at the velocity $v(p) = p$. Consequently, the initial oscillations around the right Fermi point get deformed into waves that are more and more elongated in the $x$-direction. This results in thin layers of alternating empty/filled regions along the $p$-direction. At very long times, this layered structure can be locally averaged, resulting in a smooth distribution around the R-Fermi point. Although here we formulated the argument for the hard-core limit, it is easily adapted to the finite repulsion case using Generalized Hydrodynamics~\cite{castro2016emergent,bertini_transport_2016}, see in particular~\cite{doyon2017large}.}
    \label{fig:dynamical_refermionization}
\end{figure}

Let us consider the particular case of an initial state parameterized by (in this Appendix we set the sound velocity to $v=1$)
\begin{eqnarray}
	\left< J_{\rm R}(x) \right>_{0} &=&  \frac{\sqrt{K}}{2\pi} \partial f_{\rm R}(x) , \\
	\left< J_{\rm R}(x) J_{\rm R}(y)  \right>_{0}^{\rm conn.} &=&  -\frac{ (\frac{\sqrt{K}}{2\pi} )^2}{[\frac{L}{\pi} \sin (\frac{\pi (x + i \epsilon -y)}{L} ) ]^2} ,
\end{eqnarray}
This is a state obtained from the ground state, by displacing some of the R-phonons. 
Then the fermion two-point function in the initial state is
\begin{equation}
    \label{eq:averaged}
	\langle c_{\rm R}^\dagger(x) c_{\rm R}(y) \rangle_{0} = \frac{i}{2\pi} \frac{e^{-i f_{\rm R} (x) + i f_{\rm R} (y) }}{\frac{L}{\pi} \sin (\frac{\pi (x-y + i \epsilon)}{L} )} ,
\end{equation}
and the occupation of the ${\rm R}$-fermions,
\begin{eqnarray}
	\label{eq:theta}
\nonumber	\left< c^\dagger_{{\rm R},n} c_{{\rm R},n} \right> &=& \int_0^L e^{i \frac{2\pi n}{L} v}  \left< c_{\rm R}^\dagger( v/2) c_{\rm R}( - v/2) \right> dv  .
\end{eqnarray}

\subsection{Limiting cases}
\begin{itemize}
	\item Large number of phonons: $| f_{\rm R}(x) | \gg   1$. In that case $e^{i f_{\rm R} (u+ v/2)}$ oscillates very fast, and we can evaluate the integral by the stationary phase approximation. Because of the denominator in (\ref{eq:averaged}), the integral is dominated by the neighborhood of $v=x-y=0$, and we find
	\begin{eqnarray}
	\label{eq:limclassical}
	\langle c^\dagger_{{\rm R},n} c_{{\rm R},n} \rangle & \simeq &   \frac{1}{L} \int_0^L du  \int_{-\infty}^\infty dv  \frac{i}{2\pi} \frac{e^{i [\frac{2\pi n}{L} -  \partial f_{\rm R} (u ) ] v  }}{ v +  i \epsilon }  \\
	& = &    \frac{1}{L} \int_0^L du ~\Theta(  \partial f_{\rm R} (u ) - \frac{2\pi n}{L} ) ,
\end{eqnarray}
where $\Theta(.)$ is the Heaviside step function. This is precisely what is expected from the geometric intuition illustrated in Fig.~\ref{fig:dynamical_refermionization}.

	\item Small density of phonons: $| f_{\rm R}(x) | \ll   1$. In that case we can simply expand the exponential in the numerator of (\ref{eq:averaged}),
	\begin{widetext}
		\begin{eqnarray}
			\label{eq:limquantum}
\nonumber \langle c^\dagger_{{\rm R},n} c_{{\rm R},n} \rangle & \simeq &   \frac{1}{L} \int_0^L du  \int_{0}^L dv ~  e^{i \frac{2\pi n}{L} v}  \frac{i}{2\pi} \frac{ 1 -i [ f_{\rm R} (u + v/2) - f_{\rm R} (u - v/2) ] - \frac{1}{2}  [ f_{\rm R} (u + v/2) - f_{\rm R} (u - v/2) ]^2   }{ \frac{L}{\pi} \sin (\frac{\pi v +  i \epsilon}{L} ) } \\
	& = &  \Theta(-n) - \frac{1}{2}   \int_{0}^L dv ~    \frac{i}{2\pi} \frac{  e^{i \frac{2\pi n}{L} v} }{ \frac{L}{\pi} \sin (\frac{\pi v +  i \epsilon}{L} ) }  \left( \frac{1}{L} \int_0^L   [ f_{\rm R} (u + v/2) - f_{\rm R} (u - v/2) ]^2 du \right) .
\end{eqnarray}
The first order term in the expansion does not contribute because $\int [f_{\rm R} (u + v/2) - f_{\rm R} (u - v/2)] d u = 0$.
    \end{widetext}
	
\end{itemize}

\subsection{Special case: sinusoidal wave.}
If we specialize to
\begin{equation}
	f_{\rm R}(x)  = A_0 \cos (k_0 x),
\end{equation}
corresponding to a single phonon excited w.r.t the ground state, then formula (\ref{eq:limclassical}) gives the fermion occupation
\begin{equation}
    \label{eq:arccos}
    \langle c^\dagger_{{\rm R},n} c_{{\rm R},n} \rangle \simeq  \frac{1}{\pi} {\rm arccos} \left( \frac{2\pi n/L}{A_0 k_0} \right) ,
\end{equation}
while formula (\ref{eq:limquantum}) gives
\begin{widetext}
\begin{eqnarray}
    \label{eq:quantization_phonons}
\nonumber\langle c^\dagger_{{\rm R},n} c_{{\rm R},n} \rangle & \simeq &  \Theta(-n) - \frac{1}{2}   \int_{0}^L dv ~    \frac{i}{2\pi} \frac{  e^{i \frac{2\pi n}{L} v} }{ \frac{L}{\pi} \sin (\frac{\pi v +  i \epsilon}{L} ) }   \left(  A_0^2  (1 - \cos k_0 v ) \right) \\
\nonumber	& \simeq &  \Theta(-n) 
 - \frac{A_0^2}{2}  \left( \Theta(-n) - \frac{1}{2}  \Theta(k_0-\frac{2\pi n}{L})  - \frac{1}{2}  \Theta(-k_0-\frac{2\pi n}{L})  \right) . \\
\end{eqnarray}
\end{widetext}
The result (\ref{eq:arccos}) can be interpreted semi-classically, see Fig.~\ref{fig:dynamical_refermionization}. The result (\ref{eq:quantization_phonons}), on the other hand, where one observes the emergence of plateaux as in Fig.~\ref{fig:populations}, is a consequence of phonon quantization.

\normalem
%

\end{document}